\documentclass[prl,twocolumn]{revtex4}
\usepackage{graphicx}
\def \cA{{\cal{A}} }

\def \cG{{\cal{G}} }

\def \nn{\nonumber}

\def \bk{{\bf k}}

\def \br{{\bf r}}
\def \bx{{\bf x}}

\def \dk{{\frac{d^2 k}{(2\pi)^2}}}

\begin{document}
\title{Thermoelectric transport near the pair breaking quantum phase transition out of a $d$-wave superconductor}
\author{Daniel Podolsky$^1$, Ashvin Vishwanath$^{1,2}$, Joel Moore$^{1,2}$, and Subir Sachdev$^3$}
\affiliation{$^1$Department of Physics, University of California, Berkeley, CA 94720\\
$^2$Materials Sciences Division, Lawrence Berkeley National Laboratory, Berkeley, CA 94720\\
$^3$Department of Physics, Harvard University, Cambridge, MA
02138}
\date{Printed \today}

\begin{abstract}
We study electric, thermal, and thermoelectric conductivities in
the vicinity of a $z=2$ superconductor-diffusive metal transition
in two dimensions, both in the high and low frequency limits. We
find violation of the Wiedemann-Franz law and a dc thermoelectric
conductivity $\alpha$ that does not vanish at low temperatures, in
contrast to Fermi liquids. We introduce a Langevin equation
formalism to study critical dynamics over a broad region
surrounding the quantum critical point.
\end{abstract}
\maketitle

\section{Introduction}

Transport of heat and charge provide a useful probe of strongly
correlated electronic systems. For example, a violation of the
Wiedemann-Franz law at low temperatures signals non-Fermi liquid
physics. One possible origin of such a violation is proximity to a
quantum critical point (QCP), where the presence of low energy
critical modes leads to novel phenomena. There has been much
theoretical and experimental attention directed towards electrical
transport at QCP's, in the context of the superconductor-insulator
transition in amorphous films, and the Quantum Hall transitions
(see \cite{Sachdev_book,sondhi_rmp} for reviews). Recently,
experimental investigations of thermal and thermoelectric
properties at quantum phase transitions have been pursued with
very interesting results. These include measurements of thermal
conductivity in highly underdoped cuprate systems near critical
doping \cite{bonn05,behnia_cuprate}, and of thermoelectric
transport (Nernst and Seebeck coefficients) near a heavy fermion
QCP \cite{behniaCeCoIn}. However, there has been relatively little
theoretical work on thermal and thermoelectric transport at QCP's.
In this paper we calculate the nontrivial electric, thermal, and
thermoelectric conductivities at a $z=2$, $d=2$ quantum critical
point for the transition from an unconventional (non $s$-wave)
superconductor to a diffusive metal in 2 dimensions.
%large Nernst
%signal suggests the existence of
%vortex-like excitations in the normal state of
%La$_{2-x}$Sr$_x$CuO$_4$\cite{ong}. Similarly, in the vicinity of a
%quantum phase transition, the presence of low energy collective
%modes can lead to novel transport phenomena. For instance, heat
%transport measurements at the critical doping in underdoped
%YBa$_2$Cu$_3$O$_8$\cite{bonn05} and near the critical doping in
%overdoped Tl$_2$Ba$_2$CuO$_{6+\delta}$\cite{taillefer05} show a
%large metallic heat conductivity. The linear-in-temperature
%resistivity in heavy fermion materials such as
%CeCu$_{6-x}$Au$_x$\cite{lohneysen} and
%YbRh$_2$Si$_2$\cite{steglich2000} likely originates from proximity
%to an antiferromagnetic quantum critical point. In amorphous
%films, resistivity shows scaling near the superconductor-insulator
%transition\cite{mason02}.

In an applied electric field $E$ and temperature gradient $\nabla
T$, the electric ($\sigma$), thermal ($\kappa$), and
thermoelectric ($\alpha$) conductivities are defined by,
\begin{eqnarray}
\left(\begin{array}{c} j \\ j_Q \end{array}\right)
=\left(\begin{array}{c c} \sigma & \alpha\\ \alpha T &
\tilde{\kappa} \end{array} \right) \left(\begin{array}{c} E \\
-\nabla T
\end{array}\right).
\label{eq:coeffdef}
\end{eqnarray}
where $j$ and $j_Q$ are the electrical and heat currents,
respectively.  Thermal conductivity measurements are always
carried in open circuit ($j=0$) boundary conditions, so that
$\kappa=\tilde{\kappa}-\alpha^2T/\sigma$. The difference between
$\kappa$ and $\tilde{\kappa}$ is negligible in metals, but this
need not be the case in systems with strong particle-hole symmetry
breaking, as considered here. Our results are shown in
Table~\ref{tab:results}. We find that, while the dc thermal
conductivity is metallic, the dc electric conductivity diverges as
$T\to 0$. Hence, there is a strong violation of Wiedemann-Franz
law (WF) in the bosonic sector, with an apparent {\it excess} of
charge carriers. The total conductivities are sums of fermionic
and bosonic parts, $\sigma=\sigma_f+\sigma_\Phi$ (and similarly
for $\alpha$, $\kappa$)\footnote{Maki-Thompson and density of
states corrections to the fluctuating conductivities are less
divergent than the bosonic (Aslamazov-Larkin) conductivities in
Table~\ref{tab:results}\cite{thermoelectricLargeN}}. Thus, the
magnitude of the expected WF violation is difficult to estimate.
Another striking effect is that $\alpha_\Phi$ has a very weak $T$
dependence at low temperatures ($\sim\ln\ln T$), unlike
$\alpha_f$, which vanishes linearly with $T$. Hence, experimental
measurement of the anomalous thermoelectric conductivity would
give direct information regarding critical transport properties,
as well as yielding the dimensionless ``thermoelectric figure of
merit'' $Z T = {\alpha^2 T\over \sigma \kappa}$, a measure of the
strength of particle-hole breaking.

\begin{table}[h]
\begin{center}
\begin{tabular}{|c||c|c|} \hline
    & $T=0$ & $\Omega=0$ \\ \hline
$\sigma$ & const & $\frac{(2e)^2}{h}\frac{b}{8\pi^2\eta}\ln\frac{\Lambda}{T}$ \\
$\alpha$ & $\sim i\frac{\Omega}{T}$ & $\frac{2ek_B}{h}\frac{1}{8\pi^2\eta}\ln{\ln\frac{\Lambda}{T}}$ \\
$\tilde{\kappa}/T$ &$\sim \frac{\Omega^2}{T^2}$ & const \\
\hline
\end{tabular}
\vskip0.4pc \caption{Asymptotic low $T$ behavior of transport
coefficients along the line $s=s_c$ in the zero temperature
($\Omega/T\to\infty$) and dc ($\Omega/T\to 0$) limits. The
prefactor of $\sigma(\Omega=0)$ is given in an extreme low $T$
limit $b=[{\tan^{-1} ({1}/{\eta})+{4\eta}/({1+\eta^2})}]/{2\ln\ln
(\Lambda/T)}$. Improved values for the $\Omega=0$ results, valid
for a broader $T$ range appear below Eq.~(\ref{eq:sigscaling}).
Explicit expressions for $\sigma(T=0)$ and
$\tilde{\kappa/T}(\Omega=0)$ are given in Eqs.~(\ref{sigmat0}) and
(\ref{eq:kappaom0}). \label{tab:results}}
\end{center}
\end{table}

A QCP with either Lorentz or Galilean invariance will have
infinite thermal conductivity, because a ``boosted'' thermal
distribution will never decay; the same logic applies to electric
conductivity if there is only one sign of charge carrier. We find
below that a finite $\eta$ at the $z=2$ QCP regularizes the
thermal conductivity. It is also crucial that the $z=2$ theory
breaks particle-hole symmetry so that the thermoelectric
coefficient can be nonzero; the $z=2$ theory is thus the simplest
critical theory for which all the transport coefficients in
(\ref{eq:coeffdef}) are finite.

\section{Model and currents}

Our starting point is an electron model with a pairing interaction
favoring unconventional superconductivity, and a disorder
potential whose main effect is to render electrons diffusive.
Then, introducing a Hubbard-Strata\-no\-vich field $\Phi$ in the
Cooper channel and integrating out fermions yields a
Ginsburg-Landau action for $\Phi$,\cite{herbut2000}
\begin{eqnarray}
S[\Phi]&=&\int_0^{\beta}d\tau\int
d^2r\left[\Phi^*\left(\partial_\tau+\eta
|\partial_\tau|-\frac{1}{2m}\nabla^2+s\right)\Phi\right.\nn\\
&&\left.+\frac{V}{2}|\Phi|^4\right], \label{eq:GL}
\end{eqnarray}
where the dissipative term $|\partial_\tau|$, shorthand for the
Matsubara expression $|\omega_n|$, arises from the decay of Cooper
pairs into gapless fermions. Note that, for unconventional
superconductors, disorder is pair-breaking. This insures that the
critical theory (\ref{eq:GL}) is local\cite{herbut2000}. Naive
power counting of model (\ref{eq:GL}) yields dynamical critical
exponent $z=2$. Although disorder dominates extremely close to the
transition, one can choose microscopic parameters such that
eq.~(\ref{eq:GL}) describes a large crossover region near the
QCP\cite{herbut2000}.  For instance, for a clean sample with an
elastic mean free path of 100 nm, Model (\ref{eq:GL}) is valid
provided $T$ is larger than a few
millikelvin\cite{thermoelectricLargeN}. This paper focuses on
transport in this region. Equation (\ref{eq:GL}) also applies to
the onset of antiferromagnetic order in an itinerant electron
system\cite{millis93}, with $\Phi$ an O(3) order parameter. Many
of our results apply to this QCP as well (eg. thermal
conductivity) but the field $\Phi$ carries spin, and not charge.

The electric and heat currents $j=\frac{\partial S}{\partial A_e}$
and $j_Q=\frac{\partial S}{\partial A_T}$ are obtained by making
the action (\ref{eq:GL}) gauge-covariant through the substitution
$\nabla\to{\cal D}\equiv\nabla-i e A_e-i A_T (i\partial_t)$ (see
Ref.~\cite{morenoColeman}),
\begin{eqnarray}
j&=&\frac{i e}{m}\left(\Phi^\dagger {\cal D}\Phi-({\cal D}\Phi)^\dagger\Phi\right)\nn\\
j_Q&=&\frac{1}{2m}\left((\partial_t-i s) \Phi^\dagger{\cal
D}\Phi+({\cal D}\Phi)^\dagger(\partial_t+is) \Phi\right)\nn
\end{eqnarray}
We compute conductivities from Kubo formulas involving the
dynamical correlations of these currents.

\section{Order of $T\to 0$ and $\Omega\to 0$ limits}

Conductivities near the $2d$ QCP depend on ratios of small energy
scales (and possibly a UV cutoff scale $\Lambda$), and on the
dimensionless parameters $mV$ and $\eta$,
\begin{eqnarray}
G_\gamma(\Omega)=g_\gamma\left(\frac{|s-s_c|^{\nu
z}}{T},\frac{\Omega}{T},\frac{\Lambda}{T},mV,\eta\right).
\end{eqnarray}
Here, $G_0\equiv\sigma$, $G_1\equiv\alpha$, and
$G_2\equiv\tilde{\kappa}/T$, $\Omega$ is the frequency of the
external field, and $\nu z=1$ for the $d=2$, $z=2$ model in
question. At the QCP, $s=s_c$, $G_\gamma$ depend on the order of
the limits $\Omega\to 0$ and $T\to 0$\cite{damleSachdev}. Sending
$\Omega\to 0$ first yields the dc conductivities, which are more
readily accessible to experiment, but are typically more difficult
to compute than their $T=0$ analogs.

For a $d=2$, $z=2$ theory, the quartic interaction $V$ is
dangerously irrelevant. Thus, at the $T=0$ QCP, the correct
dynamics is obtained from the limit $V\to 0^+$. We first consider
this non-interacting case. Note that, even without interactions,
finite transport is plausible, due to dissipation in
action~(\ref{eq:GL}). Indeed, $\sigma$ is metallic in the zero
temperature limit\cite{herbut2000}, as shown in the middle column
of Table~\ref{tab:results}. On the other hand, to compute dc
conductivities, we shall consider finite temperatures, for which
interactions give important logarithmic corrections.

\begin{figure}
\includegraphics[width=3 in]{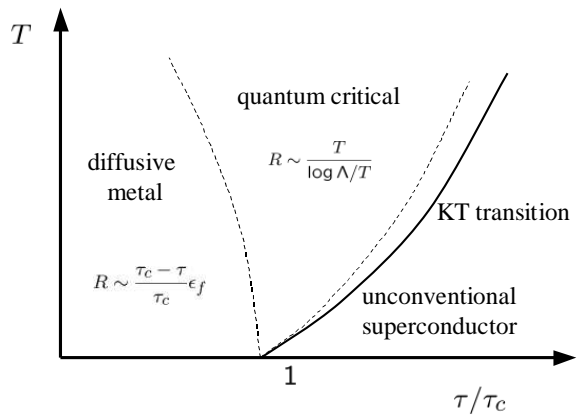}
\caption{Phase diagram in the vicinity of the QCP as a function of
disorder, parameterized by the lifetime $\tau$ of electrons in the
diffusive metal. At finite $T$, interactions shift the transition
line to the right, so that the diffusive metal phase lies above
the QCP. The crossover into the quantum critical regime occurs
when $R\approx T$.\label{fig:Rrenorm} }
\end{figure}

\section{Non-interacting case}

The conductivities are obtained from Kubo formulas,
%\begin{eqnarray}
$G_\gamma(\Omega)=\frac{(2e)^{2-\gamma}k_B^\gamma}{i\Omega
T^{\gamma}}\Pi_\gamma(\Omega+i\delta)$.
%\end{eqnarray}
For $V=0$, $\Pi_\gamma$ is given by a single-loop integral,
\begin{eqnarray}
\Pi_\gamma(i\Omega_n)&=&\frac{i^\gamma}{\beta}\sum_{\omega_n}\int\dk
\frac{k_x^2}{m^2}\left\{-\omega_n^\gamma\cG(k,\omega_n)^2\right.\label{eq:genpol0}\\
&+&\left.(\omega_n+\Omega_n/2)^\gamma\cG(k,\omega_n)\cG(k,\omega_n+\Omega_n)\right\},\nn
\end{eqnarray}
where $\cG$ is the Matsubara Green's functions,
$\cG(k,\omega_n)=\left(i\omega_n-\eta|\omega_n|-\epsilon_k-s\right)^{-1},$
where $\epsilon_k=k^2/2m$.  Rewriting this in terms of the
spectral function,
$\cG(k,\omega_n)=\int\frac{d\omega'}{2\pi}\frac{\cA(k,\omega')}{i\omega_n-\omega'}$,
$\cA(k,\omega)=\frac{2\eta\omega}{(\omega-\epsilon_k-s)^2+\eta^2\omega^2}$,
we obtain,
\begin{eqnarray}
\Pi_\gamma(i\Omega_n)&=&\frac{1}{\beta}\sum_{\omega_n}\int_{\omega_1,\omega_2,\bk}
\left(\frac{\omega_1+\omega_2}{2}\right)^\gamma
\cA(k,\omega_1)\cA(k,\omega_2)\nn\\
&\,&\frac{k_x^2}{m^2}\left\{\frac{1}{(i\omega_n-\omega_1)(i(\omega_n+\Omega_n)-\omega_2)}\right.\nn\\
&\,&\left.-\frac{1}{(i\omega_n-\omega_1)(i\omega_n-\omega_2)}\right\}\label{eq:genpol},
\end{eqnarray}
%where
%\begin{eqnarray}
%$\cA(k,\omega)=\frac{2\eta\omega}{(\omega-\epsilon_k-s)^2+\eta^2\omega^2}$ %\nn
%\end{eqnarray}
%is the spectral function.
In (\ref{eq:genpol}) we have made the
substitution $i\omega_n\to(\omega_1+\omega_2)/2$. This accounts
for the fact that the time derivative in $j_Q$ does not commute
with the time-ordering symbol, and is necessary to obtain the
correct thermal conductivity~\cite{ambegaokarGriffin}. Performing
the $\omega_n$ sum and using the identity,
$f^0(\omega+i\Omega_n)=f^0(\omega)$, before analytically
continuing to real frequencies, $i\Omega_n\to\Omega+i\delta$, we
find
\begin{eqnarray}
{\rm Im}&&\Pi_\gamma(\Omega+i\delta)
=\frac{1}{2}\int_{\omega_2,\bk}
\frac{k_x^2}{m^2}\left(\omega_2-\frac{\Omega}{2}\right)^\gamma\times\label{ambegPol}\\
&&\cA(k,\omega_2-\Omega)\cA(k,\omega_2)\left(f^0(\omega_2-\Omega)-f^0(\omega_2)\right),\nn
\end{eqnarray}
where $f^0(\omega)=(\exp\left(\beta\omega\right)-1)^{-1}$ is the
Bose occupation factor. The integrals are UV finite, and hence the
dependence on the cutoff $\Lambda$ drops out,
\begin{eqnarray}
G_\gamma^{\rm scaling}(\Omega)=g^{\rm
scaling}_\gamma\left(\frac{|s-s_c|^{\nu
z}}{T},\frac{\Omega}{T},\eta\right).\nn
\end{eqnarray}
Expression~(\ref{ambegPol}) can readily be evaluated for generic
$\Omega/T$.

\subsection{Zero temperature ($\Omega/T\to \infty$) limit}
The results in the zero
temperature limit at the QCP ($s=s_c=0$) are summarized in the
middle column of Table~\ref{tab:results}. While $\sigma$ is
metallic, $\alpha$ and $\tilde{\kappa}/T$ are divergent (note that
$\alpha$ is purely reactive). The electric conductivity depends on
the dissipation $\eta$,
\begin{eqnarray}
\sigma\left(\frac{\Omega}{T}\to\infty\right)&=&\frac{(2e)^2/h}{4\pi
\eta^2
}\left\{[2\eta+\pi(1+\eta^2)]\arctan\eta\right.\nn\\
&\,&\left.-\pi\eta-\eta^2-(1+\eta^2)\arctan^2\eta\right\},
\label{sigmat0}
\end{eqnarray}
since $\eta$ is marginal.

\subsection{DC ($\Omega/T\to 0$) limit}
On the other hand, the dc transport properties at $V=0$ are
drastically different. Whereas
$\sigma=\frac{(2e)^2}{h}\frac{T}{4|\Omega|}$ and
$\alpha=\frac{2ek_B}{h}\frac{1}{8\pi^2\eta}\ln\frac{T}{|\Omega|}$
are divergent, $\frac{\tilde{\kappa}}{T}$ is metallic,
\begin{eqnarray}
\frac{\tilde{\kappa}}{T}\left(\frac{\Omega}{T}\to
0\right)=\frac{k_B^2}{h}\left(\frac{\eta+\arctan
(1/\eta)}{12\pi\eta}\right) \label{eq:kappaom0}
\end{eqnarray}
In order to elucidate the dc results, we employ a simple Boltzmann
equation approach, which is exact for weak dissipation $\eta\to
0$. The dissipating term in the action (\ref{eq:GL}) can be
interpreted as the self-energy of bosons due to their interactions
with the fermion bath,
\begin{eqnarray}
\Sigma(i\omega_n)=\eta|\omega_n|.\nn
\end{eqnarray}
This is purely imaginary for real frequencies ($|z|=z\,{\rm
sign}({\rm Re}\,z)$). Hence, while the energy of a quasi particle
is not renormalized, quasi particles are given a finite lifetime,
\begin{eqnarray}
\frac{1}{\tau_k}&=&i\left[\Sigma(\omega+i\delta)-\Sigma(\omega-i\delta)\right]_{\omega=\epsilon_k}=2\eta\epsilon_k
\label{tau}. \label{eq:tau}
\end{eqnarray}
This is the boson lifetime, at bubble level, in the fermion model.
Here we neglect corrections to the transport lifetime from higher
order diagrams.

The Boltzmann equation for the occupation function $f(\bk,\br,t)$
in the scattering lifetime approximation yields
\begin{eqnarray}
\partial_t f+{\bf v}\cdot \nabla_\br f+\dot{\bk}\cdot\nabla_\bk
f=-\frac{1}{\tau_\bk}(f-f^0).
\end{eqnarray}
The electric and energy currents are expressed in terms of $f$,
\begin{eqnarray}
j^a(\br,t)&=&\int\dk \,(2e) v_\bk^a f(\bk,\br,t)\\
j^a_Q(\br,t)&=&\int\dk \,(\epsilon_\bk-s) v_\bk^a f(\bk,\br,t).
\end{eqnarray}
The conductivities are then
\begin{eqnarray}
G_\gamma=\frac{(2e)^{2-\gamma}k_B^\gamma}{T^\gamma
\hbar}\int\dk\,\frac{\tau_\bk}{1-i\Omega\tau_\bk}
\frac{\epsilon_k^{1+\gamma}}{m}\left(-\frac{\partial
f^0}{\partial\epsilon_\bk}\right).\label{eq:boltzcond}
\end{eqnarray}
For $\eta\to 0$, this gives perfect agreement with the previous
results for $G_\gamma$ in the dc limit. Here, we see that the
reason for the divergences in $\sigma$ and $\alpha$ is that low
energy quasiparticles have arbitrarily large lifetimes $\tau_\bk$.
On the other hand, these very long-lived quasiparticles, having
low energies, contribute little to the energy current. Hence
$\tilde{\kappa}/T$ is finite.

\section{Interacting case}

For $T>0$, the dangerously irrelevant interaction $V$ must be
taken into account.  The most important effect of interactions is
to shift the phase transition, such that the QCP is approached at
finite temperatures from the diffusive metal phase, as shown in
Fig.~\ref{fig:Rrenorm}. This is captured by a renormalized mass
$R$, discussed below, which is positive above the transition.  $R$
acts as an effective gap for low energy quasiparticles, thus
rendering all dc transport coefficients finite. The situation is
in contrast with Ref.~\cite{damleSachdev}, where interactions
regularize transport by introducing quasiparticle scattering.
Here, the leading low $T$ dc conductivities are obtained from a
Hartree-Fock (HF) analysis, where the only effect of interactions
is to shift the quasiparticle mass $R$.  The HF results are shown
in the last column of Table \ref{tab:results}. These results are
valid for extremely low temperatures, such that
$\ln\ln\Lambda/T\gg 1$. To study transport on a much broader
region surrounding the QCP, we introduce a classical treatment of
the order parameter which, when supplemented by Langevin dynamics,
will be shown to capture the correct quantum critical transport
behavior over the region where the much weaker condition,
$\ln\Lambda/T\gg 1$, is satisfied.

\subsection{Classical action for order parameter}

With interactions, the critical point is shifted away from
$s_c=0$. To linear order in $V$,
\begin{eqnarray}
s_c=2V\int_{\omega,k}\frac{1}{i\omega_n-\eta|\omega_n|-\epsilon_k}\nn
%\cG(k,\omega)
\end{eqnarray}
In the vicinity of the QCP, we renormalize $V$ by a one loop RG
equation, up to the scale where the system either develops a gap,
or when the rescaled temperature reaches an upper frequency cutoff
$\Lambda_\omega$\cite{fisherhohenberg,millis93},
\begin{eqnarray}
V_R\approx\frac{2\pi^2}{m\left(\tan^{-1}\frac{1}{\eta}+
\frac{4\eta}{1+\eta^2}\right)}\frac{1}{\ln\frac{\Lambda_\omega}{{\rm
Max}(T,|s-s_c|)}}\nn
\end{eqnarray}
The static properties of the finite $T$ model can be studied by
integrating out all non-zero Matsubara frequency modes. After
rescaling,
%\begin{eqnarray}
$\Phi=\sqrt{2mT}\psi$, %\nn\\
$\br=\bx/\sqrt{2m},$ %\nn
%\end{eqnarray}
the $\omega_n=0$ mode has the following classical action,
\begin{eqnarray}
S_c=\int d^2 x\left[|\nabla_x
\psi|^2+\tilde{R}|\psi|^2+\frac{U}{2}|\psi|^4\right]\nn
\end{eqnarray}
where $U=2mTV_R$. This theory is super-renormalizable, and is
rendered UV finite by introducing a renormalized mass $R$,
\begin{eqnarray}
\tilde{R}=R-2U\int_0^\Lambda \dk \frac{1}{k^2+R}.
\end{eqnarray}
$R$ has a universal expression in terms of $s-s_c$, reflecting the
contribution of the $\omega_n \neq 0$ modes
\begin{eqnarray}
R&=&(s-s_c)+\frac{U}{2\pi
T}\int_0^{\infty}dy\left[\frac{T}{y+R}-\frac{T}{y+s-s_c}
\right.\nn\\
&\,&\left.+\int_0^\infty\frac{d\Omega}{\pi}\frac{\eta\Omega}{e^{\Omega/T}-1}\left(\frac{1}{(\Omega-y-(s-s_c))^2+\eta^2\Omega^2}\right.\right.\nn\\
&\,&+\left.\left.\frac{1}{(\Omega+y+s-s_c)^2+\eta^2\Omega^2}\right)\right]\nn\\
&=&s-s_c+\frac{U}{2\pi}\left\{\ln\frac{T}{R}+
F(s-s_c,\eta)\right\},\label{eq:selfcons}
\end{eqnarray}
where $F(s-s_c,\eta)\to \ln\sqrt{1+\eta^2}$ as $s\to s_c$. Solving
this self-consistent equation at $s=s_c$ yields,
\begin{eqnarray}
R\sim \frac{T}{\ln(\Lambda_\omega/T)}\label{Rcrit}
\end{eqnarray}
up to a prefactor of order $\ln\ln(\Lambda_\omega/T)$. Note that,
for $T>0$, $R$ is always positive, even for arbitrarily negative
values of $s-s_c$. This is due to the absence of long-range order
(LRO) in $2d$ at finite $T$. For an O(2) order parameter, however,
quasi LRO is established at a $T>0$ Kosterlitz-Thouless
transition.

A description in terms of a classical action \cite{dunkel} is
appropriate whenever $\log\Lambda_\omega/T \gg 1$. In this limit,
$U \ll T$, so that modes with $\omega_n\ne 0$ are significantly
gapped.

\subsection{Dynamics of order parameter}

We approximate the low frequency dynamics of the classical order
parameter by a Langevin equation (model A dynamics of
Ref.~\cite{hohenberg_halperin}),
\begin{eqnarray}
\frac{\partial\psi}{\partial t}&=&-(i+\eta)\frac{\delta
S_c}{\delta
\psi^*}+f_\eta\label{eq:langevin}\\
\langle f_\eta^*(x,t)
f_\eta(x',t')\rangle&=&2\eta\delta(x-x')\delta(t-t')\nn
\end{eqnarray}
Equal time correlators computed with these dynamics are equal to
those of the classical action $S_c$, as necessary. The appearance
of the ``bare" value of $\eta$ in eq.~(\ref{eq:langevin}) is due
to the fact that dispersion in the quantum action is non-local in
time and therefore is not renormalized.

Consider the HF approximation, in which (\ref{eq:langevin})
becomes a linear equation with mass $R$. Solving for $\sigma$,
%\begin{eqnarray}
%\sigma(\Omega)=\frac{(2e)^2}{h}\frac{T}{2\pi i\Omega}
%\ln\left(1+\frac{i\Omega}{4\eta R}\right)\nn
%\end{eqnarray}
%The real part of $\sigma$ is then
\begin{eqnarray}
{\rm Re}\,\sigma(\Omega)&=&\frac{(2e)^2}{h}
\frac{T}{2\pi\Omega}\tan^{-1}\frac{\Omega}{4\eta R}\nn\\
&\to&\frac{(2e)^2}{h}\left\{
\begin{array}{c l}
\frac{T}{4|\Omega|} & {\rm for }\,R=0 \\
\frac{T}{8\pi \eta R} & {\rm for } R\ne 0,\,\Omega\to 0
\end{array}\right.\nn
\end{eqnarray}
For $R=0$, this reproduces the non-interacting result, as
expected. On the other hand, for $R\ne 0$, we obtain a finite dc
conductivity,
\begin{eqnarray}
\sigma\sim \ln\frac{\Lambda_\omega}{T} \label{eq:sigma}
\end{eqnarray}
We note that Eq.~(\ref{eq:sigma}) disagrees with Ref.~\cite{ddpp},
which predicts $\sigma\sim \ln\ln(\Lambda_\omega/T)$. This is due
to an erroneous computation of $T^*$ in Eq.~(12) of that
reference.

Naive use of Eq.~(\ref{eq:langevin}) yields divergent values of
$\alpha$ and $\tilde{\kappa}/T$. This is not surprising: the
Langevin equation assumes classical modes, whose occupation
factors satisfy equipartition, $f_{\rm eq}(\omega)=T/\omega$.
However, inspection of the Boltzmann approach,
eq.~(\ref{eq:boltzcond}), shows that for such distribution,
$\alpha$ and $\tilde{\kappa}/T$ have UV catastrophes. In this
sense, the Langevin equation does not capture the correct dynamics
of high energy quantum modes. However, high energy modes are very
weakly perturbed by the quartic interaction. Thus, the correct
result is given by the Boltzmann equation with a full Bose
distribution, as in Eq.~(\ref{eq:boltzcond}), but with a chemical
potential set by $R$. Equivalently, this corresponds to evaluating
the one-loop quantum expression (\ref{eq:genpol}) with chemical
potential $R$. This yields the last column of
Table~\ref{tab:results}.

\begin{figure}
\includegraphics[width=3 in]{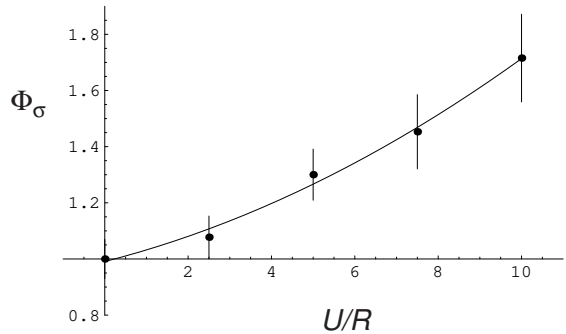}
\caption{Estimated scaling function $\Phi_\sigma(U/R,\eta=1)$ from
numerical integration of the Langevin equation
(\ref{eq:langevin}). Convergence for static quantities of the
Langevin algorithm was tested by comparison with results from a
Wolff cluster algorithm. Each data point represents at least $2
\times 10^7$ time steps on a lattice of spacing $a=0.1$ and linear
size $L=64$, which for static quantities with $\eta=0$ is known to
approximate well the continuum limit $a \rightarrow 0$. Results
were normalized by the $U/R=0$ result and fit by a quadratic
polynomial (solid line). \label{fig:lfig} }
\end{figure}

The HF results can be obtained independently from an exact
solution of the quantum model~(\ref{eq:GL}) in the large $N$
limit\cite{thermoelectricLargeN}, where $N$ is the number of
components of the order parameter ($N=2$ for superconductivity).
This is an important check that the Langevin equation
(\ref{eq:langevin}) captures the correct universal dynamics. We
see from Eq.~(\ref{eq:selfcons}) that, for $s=s_c$ and at low $T,$
\begin{eqnarray}
\frac{U}{R}\propto\frac{1}{\ln T/R}\sim\frac{1}{\ln\ln
\Lambda_\omega/T}\nn
\end{eqnarray}
which justifies HF provided that $\ln\ln (\Lambda_\omega/T)$ is a
large number. More generally, to go beyond HF, we must consider
higher order corrections in $U$. From the Kubo formula and the
fluctuation-dissipation theorem, we deduce that the dc electric
conductivity obeys
\begin{eqnarray}
\sigma=\frac{(2e)^2}{h}\frac{T}{8\pi\eta
R}\Phi_\sigma\left(\frac{U}{R},\eta\right),\label{eq:sigscaling}
\end{eqnarray}
for some scaling function $\Phi_\sigma$ satisfying
$\Phi_\sigma(0,\eta)=1$. A similar analysis applies to $\alpha$
and $\tilde{\kappa}/T$, with the important difference that
substractions are necessary to cancel leading UV divergences, as
discussed above. When working with the renormalized $R$, the only
UV divergence comes from the diagrams already computed. Thus,
\begin{eqnarray}
\alpha=\alpha_{\rm quantum, 1\,
loop}+\frac{2e}{2}\Phi_\alpha\left(\frac{U}{R},\eta\right),\nn\\
\frac{\tilde{\kappa}}{T}=\frac{\tilde{\kappa}_{\rm quantum, 1\,
loop}}{T}+\frac{R}{2T}\Phi_\kappa\left(\frac{U}{R},\eta\right),\nn
\end{eqnarray}
where the small $R$ limit of the (quantum, 1 loop) results is
$\frac{2 e k_B}{8 \pi^2 \eta h} \ln\frac{T}{R}$ for $\alpha$, and
Eq.~(\ref{eq:kappaom0}) for $\tilde{\kappa}/T$. The functions
$\Phi_\gamma$ can be evaluated numerically by introducing a
lattice,
\begin{eqnarray}
S_{cL}=\sum_{\langle
ij\rangle}|\psi_i-\psi_j|^2+\sum_i\left[\tilde{R}_La^2|\psi_i|^2+\frac{Ua^2}{2}|\psi_i|^4\right]\nn
\end{eqnarray}
and requiring that the renormalized $R$ be the same in the lattice
and continuum theories,
\begin{eqnarray}
\tilde{R}_L=R-2U\int_{-\pi}^{\pi}\frac{dk_x}{2\pi}\int_{-\pi}^{\pi}\frac{dk_y}{2\pi}
\frac{1}{4-2\cos k_x-2\cos k_y+Ra^2}.\nn
\end{eqnarray}
Fig.~\ref{fig:lfig} shows the scaling function
$\Phi_\sigma(U/R,\eta=1).$ This, combined with
Eqs.~(\ref{eq:selfcons}) and~(\ref{eq:sigscaling}), gives the
electric conductivity for the entire quantum critical regime. We
stress that these results rely only on the condition
$\log\Lambda/T \gg 1$. This use of Langevin dynamics to obtain
full scaling functions for transport quantities at a QCP should be
applicable at many other transitions. Here, we have used them to
find WF violation and an anomalous thermoelectric conductivity at
a transition of experimental interest.

\section{Conclusions}

\section{Acknowledgments}
We benefited from useful discussions with B. Binz, A. Paramekanti,
T. Senthil, and members of the 2005 Aspen Center for Physics
Workshop on Competing Orders, where part of this work was
completed. This work was supported by NSF grants DMR-0238760
(J.M.) and DMR-0537077 (S.S.), the Hellman Fund (J.M.), and the
LDRD program of LBNL under DOE grant DE-AC02-05CH11231 (D.P. and
A.V.).

%Main effect of U: not to scatter, but to shift R

%\nocite{}
%\bibliography{d:/Research/bibs}

%\end{document}

\end{document}